\documentclass[aip, rsi, amsmath, amssymb, groupaddress, reprint]{revtex4-1}

\usepackage{graphicx}
\usepackage{dcolumn}
\usepackage{bm}

\usepackage[version-1-compatibility,separate-uncertainty=true]{siunitx}

\usepackage{hyperref}
\usepackage{color}

\begin{document}


\title[]{Compact setup for the production of  $^{87}$Rb $|F=2, m_{F}=+2\rangle$ Bose-Einstein condensates in a hybrid trap}

\author{Raffaele Nolli}
\author{Michela Venturelli}
\author{Luca Marmugi}
\email{l.marmugi@ucl.ac.uk}
\author{Arne Wickenbrock}
\altaffiliation[Current address: ]{Johannes Gutenberg-Universit\"at Mainz, 55128 Mainz, Germany.}
\author{Ferruccio Renzoni}
\affiliation{Department of Physics and Astronomy, University College London, Gower Street, London WC1E 6BT, United Kingdom}

\date{\today}

\begin{abstract}
We present a compact experimental apparatus for Bose-Einstein condensation of $^{87}$Rb in the $|F=2, m_{F}=+2\rangle$ state. A pre-cooled atomic beam of ${}^{87}$Rb is obtained by using an unbalanced magneto-optical trap, allowing controlled transfer of trapped atoms from the first vacuum chamber to the science chamber. Here, atoms are transferred to a hybrid trap, as produced by overlapping a magnetic quadrupole trap with a far-detuned optical trap with crossed beam configuration, where forced radiofrequency evaporation is realized.  The final evaporation leading to Bose-Einstein condensation is then performed by exponentially lowering the optical trap depth. Control and stabilization systems of the optical trap beams are discussed in detail. The setup reliably produces a pure condensate in the $|F=2, m_{F}=+2\rangle$ state in $\SI{50}{\second}$, which include $\SI{33}{\second}$ loading of the science magneto-optical trap and $\SI{17}{\second}$ forced evaporation.
\end{abstract}

\pacs{37.10.De, 37.10.Gh, 67.85.-d, 67.85.Hj}
                             
\keywords{Bose-Einstein Condensation, Low-Velocity Intense Source, Far-detuned Optical Dipole Trap \\
\vskip 8pt
\centering \emph{This article may be downloaded for personal use only. Any other use requires prior permission of the authors and AIP Publishing. The article appeared in Rev.~Sci.~Instr.~\textbf{87}, 083102 (2016) and may be found at \href{http://dx.doi.org/10.1063/1.4960395}{http://dx.doi.org/10.1063/1.4960395}.}}

\maketitle

\section{Introduction}\label{sec:intro}
Since their first realizations \cite{anderson1995, davis1995}, ultra-cold quantum gases of dilute alkali vapors have played a central role in atomic physics, with tremendous impact in a number of fields, ranging from atom interferometry \cite{shin2004, maussang2010}, to quantum phase transitions \cite{greiner2002, spielman2007, valtolina2015} and quantum simulations \cite{bloch2012, aidelsburger2013, banerjee2013, miyake2013}, just to name a few examples.

Although many different approaches and technical solutions have been demonstrated \cite{altin2010, paris2014, ivory2014}, all implementations rely on laser cooling and forced evaporative cooling. In this context, red-detuned optical dipole traps  \cite{grimm1999} are nowadays widely used for a fast and robust path to quantum degeneracy \cite{lin2009, burchianti2014, mishra2015}. Their success is mainly due to the large optical access, the high level of experimental control and flexibility, and the possibility to achieve state-insensitive trapping.

There is still great interest in Bose-Einstein condensates (BECs) of ${}^{87}$Rb as robust tools for simulation of quantum phenomena \cite{clement2009, lin2011, schafer2014, eto2014, hamner2015, bagnato02, bagnato01, grossert2016}. Hence, we report here on a compact setup for the realization of an ${}^{87}$Rb BEC in the $|F=2, m_{F}=+2\rangle$ state.  Quantum degeneracy is achieved in a hybrid optical crossed dipole trap (OCDT). The atomic sample is obtained from vapor phase by means of a low velocity intense source (LVIS) \cite{lvis}. A first forced evaporation is performed in a magnetic quadrupole trap (MQT); then the atoms are transferred to the red-detuned OCDT, where condensation is achieved. A pure BEC of $\sim 5 \times 10^{4}$ ${}^{87}$Rb atoms in the $|F=2, m_{F}=+2\rangle$ state is obtained at a temperature around $\SI{10}{\nano\kelvin}$ every $\SI{50}{\second}$, with a $\SI{17}{\second}$ evaporation to a pure quantum gas.

Here, the implementation of the LVIS arrangement eliminates the need for a more complicated, bulkier or noisier transport setup. The hybrid trap, as constituted by a MQT and an OCDT, allows for a larger optical access to the trapping area with respect to pure magnetic traps and for a reduction of the laser intensity with respect to a pure optical trap, without the need of a push beam to compensate the gravitational force.

The paper is organized as follows: in Sec.~\ref{sec:hight} the LVIS and the experimental protocol for pre-cooling in the MQT are described. In Sec.~\ref{sec:lowt}, transfer to the hybrid MQT/OCDT and the forced evaporation sequence are described. In Sec.~\ref{subsec:dt}, we describe the setup of the OCDT and in Sec.~\ref{subsec:dtchar} we present measurements to test the OCDT. Finally, in Sec.~\ref{sec:exp} an experimental characterization of the BEC is reported.

\section{Low Velocity Intense Source and Magnetic Quadrupole Trap}\label{sec:hight}
\begin{figure*}[htbp]
\centering
\includegraphics[width=0.55\linewidth]{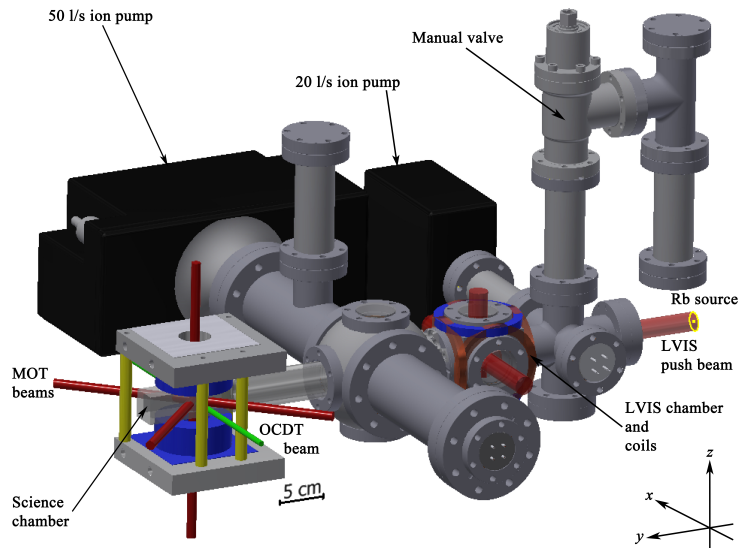}
\caption{Sketch of the vacuum chambers for ${}^{87}$Rb BEC. The Rb solid reservoir is contained at the bottom of the blind flange on the right hand side. LVIS and MOT main coils are high-lighted in blue; respective laser beams are indicated in red. Green labels the optical dipole trap beam. For the sake of simplicity, only one beam of the optical crossed dipole trap is shown.}\label{fig:chamber}
\end{figure*}

The vacuum system is sketched in Fig.~\ref{fig:chamber}.

Atomic Rb vapor is obtained from a solid reservoir, contained in a sealable arm of the vacuum system, terminated by a blind flange (right hand side of of Fig.~\ref{fig:chamber}). The reservoir is kept at a constant temperature of about $\SI{310}{\kelvin}$ by means of resistive heating. The vapor fills the first chamber, a standard six way cube steel chamber. It is evacuated by a $\SI{20}{\l  \second^{-1}}$ ion pump (Varian VacIon Plus 20 StarCell).

Here, atoms are collected in a 3D magneto optical trap (MOT), in the so-called LVIS configuration \cite{lvis}: the cooling beam oriented along $-\hat{y}$ (see Fig.~\ref{fig:chamber}) is obtained by retroreflection from a dielectric 1'' mirror, equipped with a quarter-wave plate, glued at the leftmost flange of the LVIS chamber. A hole of $\SI{1.5}{\milli\meter}$ is drilled in the center of the optics, to create a `dark' column in the center of the trapped atomic cloud. In this way, the unbalanced radiation pressure along $\hat{y}$ pushes a collimated, pre-cooled and isotopically selected atomic beam of ${}^{87}$Rb into the science chamber (left-hand side in Fig.~\ref{fig:chamber}).

The LVIS chamber and the science chamber are connected by a spherical square connector; the total distance between the centers of the two chambers is $\SI{330}{\milli \meter}$. The spherical connector is connected to a second ion pump, with $\SI{50}{\l \second^{-1}}$ capacity (Varian VacIon Plus 50 StarCell). Here, a SAES getter pump is installed, but not currently in use. The science chamber is a glass cubic cell, sized $\SI{30}{\milli\meter} \times \SI{30}{\milli\meter} \times \SI{100}{\milli \meter}$. The residual background pressure is $\SI{<1e-10}{\milli \bar}$.

In the science chamber, a second 3D MOT is created, with direct loading from the LVIS, at a rate of $\SI{19.5 \pm 0.3 }{\times ~10^{6}~ \second^{-1}}$. A pair of anti-Helmholtz coils supplies the required magnetic field gradient. Coils are made of $184$ turns of copper enamelled wire (diameter $\SI{1}{\milli \meter}$); the minimum on-axis width (inner diameter) of the coils is $\SI{23}{\milli \meter}$, while the external diameter is $\SI{62}{\milli \meter}$. The coils are water cooled and the current is actively stabilised by means of a Honeywell CSNP661 current sensor and a custom designed MOSFET electronic feedback servo. Three pairs of mutually orthogonal compensation coils allow zeroing of the local DC magnetic field.

The cooling beams, for both the LVIS and the MOT, are supplied by a master oscillator power amplifier (MOPA) setup. A Radiant Dyes Narrowdiode external cavity diode laser (ECDL) serves as the seed for a home-built amplifier system based on a M2K-TA-0780-2000 tapered amplifier. Frequency stabilization is obtained by means of frequency modulation lock, using the saturated absorption spectroscopy of Rb vapor as a reference. The output is fed to a single-mode polarization maintaining optical fiber, which delivers light to the experimental chambers and acts as a spatial filter. At the science chamber, each cooling beam has a $1/e$ diameter of $\SI{25}{\milli \meter}$ and a power of $\SI{5}{\milli\watt}$ ($\sim I_{sat}\approx\SI{1.67}{\milli \watt \centi \meter^{-2}}$).

The repumping beams are provided by a Moglabs CEL002 ECDL, with a total output of $\SI{70}{\milli\watt}$. In this case, frequency is stabilized by means of dichroic atomic vapor laser lock  (DAVLL)\cite{davll}, and the laser output is split and routed to the LVIS and the science chamber. 

Computer-controlled acousto-optic modulators (AOMs) and mechanical shutters allow independent tuning and intensity control of all the beams during the experimental sequence. The apparatus is controlled and continuously monitored by a Labview program.

\subsection{Experimental protocol for pre-cooling in the magnetic quadrupole trap}
\begin{figure*}[htbp]
\centering
\includegraphics[width=0.7\linewidth]{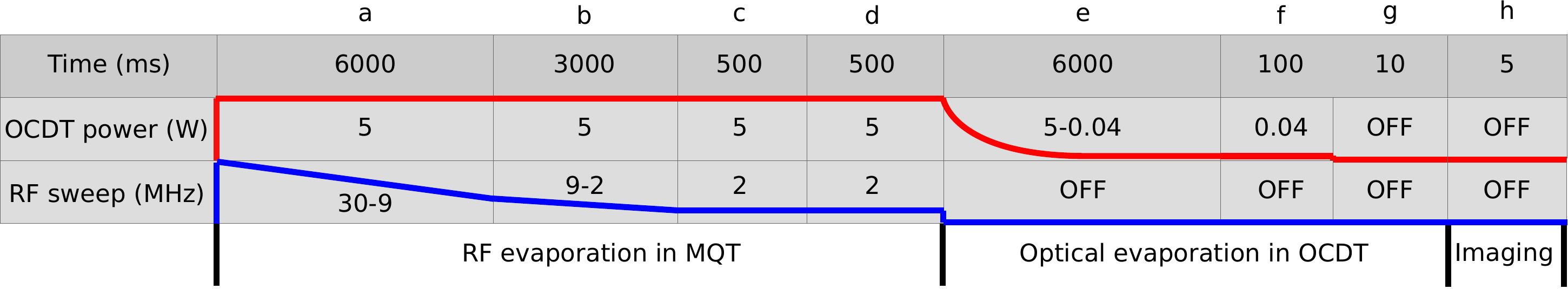}
\caption{Experimental sequence for generation of a BEC of ${}^{87}$Rb in a hybrid trap consisting of an OCDT and an MQT.} \label{fig:seq}
\end{figure*}

During the MOT loading in the science chamber, the cooling laser is detuned by $-0.96\Gamma$ with respect to the ${}^{87}$Rb $D_{2}$ line $F = 2 \rightarrow F' = 3$ cycling transition, where $\Gamma = 2\pi \cdot 6.066 \si{~\mega \hertz}$ is the natural linewidth. The magnetic field gradient is stabilized at $\SI{24}{\gauss \centi \meter^{-1}}$. After $\SI{30}{\second}$ of loading, more than $3 \times 10^{8}$ atoms are collected in the MOT with a temperature $T_{MOT}=\SI{408 \pm 20}{\micro \kelvin}$, where the uncertainty, as in the following, is the standard deviation of different independent measurements. The $1/e$ trap lifetime, in these conditions limited mainly by the MOT internal dynamics, is $\SI{51.6 \pm 0.2}{\second}$.

The atomic density is increased during a $\SI{5}{\milli \second}$ compression phase, in which the laser detuning is set to $-1.5\Gamma$ and the magnetic field gradient to $\SI{30}{\gauss \centi \meter^{-1}}$. After compression, the atoms undergo $\SI{8}{\milli \second}$ of polarization gradient cooling; for this, the magnetic field gradient is switched off and the trap laser detuning is further increased to $-3.4\Gamma$. With this method, temperatures as low as $\SI{40 \pm 1}{\micro \kelvin}$ and densities $> 4 \times 10^{10} \si{\per \cubic \centi \meter}$ are obtained.

The atoms are then optically pumped for $\SI{0.2}{\milli \second}$ to the low field-seeking state $|F = 2, m_F = +2\rangle$, which allows loading of more than $1.8 \times 10^{8}$ atoms in the MQT. Duration and detuning ($-1.8\Gamma$) of the optical pump pulse are empirically optimized in order to ensure the most efficient transfer to the MQT and minimum heating of the atomic sample. It is noteworthy that, at this stage of the experimental sequence, trapping in the  $|F = 1\rangle$ states could be easily achieved by performing a temporal dark MOT \cite{darkmot01, darkmot02}, prior to the polarization gradient cooling.

Optical pumping is realized by the cooling beam aligned along $-\hat{z}$.  In order to ensure fast ($\ll \SI{1}{\milli \second}$) and stable switch off of the magnetic field gradient and switch on of the uniform bias field for optical pumping (along $-\hat{z}$), an extra pair of Helmholtz coils is installed, coaxially with the main anti-Helmholtz coils and the $\hat{z}$ compensation coils' pair (not shown in Fig.~\ref{fig:chamber}). The idea is to overcompensate the background DC magnetic field along the main coils' axis by means of the compensation coils, and to zero this extra component with the uniform field generated by the Helmholtz coils ($\mathbf{B_{H}}$).

During the $\SI{0.2}{\milli \second}$ optical pumping phase, $\mathbf{B_{H}}$ is switched off. As a result, a uniform magnetic field produced by the compensation coils is obtained along $-\hat{z}$. After the optical pumping phase, $\mathbf{B_{H}}$ is switched on again, thus allowing exact zeroing of the local DC fields.

At the same time, the magnetic field gradient is switched on at $\SI{112}{\gauss \centi \meter^{-1}}$ for $\SI{50}{\milli \second}$ to allow loading of the MQT. Then, the gradient is linearly ramped up to $\SI{225}{\gauss \centi \meter^{-1}}$ in $\SI{150}{\milli \second}$ to increase the spatial density of the atomic cloud. The ramp rate is a critical parameter for ensuring rapid compression and thus limiting heating of the atoms.

\section{Optical Crossed Dipole Trap and Final Evaporation}\label{sec:lowt}
At this stage, the OCDT is loaded. The relevant experimental sequence for loading the OCDT is summarized in Fig.~\ref{fig:seq}, as well as the final stages of the protocol, leading to BEC.

After the MQT is loaded, radio-frequency (RF) forced evaporative cooling is applied, directly driving transitions between the $|F = 2, m_F = +2\rangle$ state to untrapped states. RF is supplied by a circular copper coil of diameter $\SI{30}{\milli \meter}$, whose axis coincides with that of the anti-Helmholtz coils. The RF power is kept constant at $\SI{14}{\decibel m}$ ($\SI{\approx25}{\milli \watt}$).

The RF evaporation is composed of two frequency sweeps, controlled independently via Labview. The first one lasts  $\SI{6}{\second}$; frequency is linearly swept from $\SI{30}{\mega \hertz}$ to $\SI{9}{\mega \hertz}$, with a constant MQT field gradient of $\SI{225}{\gauss \centi \meter^{-1}}$ (\textbf{a} in Fig.~\ref{fig:seq}). 

The second sweep lasts $\SI{3}{\second}$, and the frequency is linearly reduced from $\SI{9}{\mega \hertz}$ to $\SI{2}{\mega \hertz}$. At the same time, the MQT field gradient is linearly ramped down to $\SI{30}{\gauss \centi \meter^{-1}}$ (\textbf{b} in Fig.~\ref{fig:seq}). The RF is then kept at $\SI{2}{\mega \hertz}$ for $\SI{500}{\milli \second}$ with a MQT gradient of $\SI{30}{\gauss \centi \meter^{-1}}$ (\textbf{c} in Fig.~\ref{fig:seq}). A decompression is then performed for $\SI{500}{\milli \second}$, with a gradient's linear decrease to $\SI{20}{\gauss \centi \meter^{-1}}$ (\textbf{d} in Fig.~\ref{fig:seq}). After that, the RF is switched off.

As a result, $3 \times 10^{6}$ atoms are loaded in the OCDT, with an average temperature of $\SI{5.0\pm 0.6}{\micro \kelvin}$. A background magnetic field gradient of $\SI{20}{\gauss \centi \meter^{-1}}$ is maintained to aid levitation of the atomic cloud during the following evaporation.

During the OCDT loading, the minima of the optical and magnetic traps are displaced by $\SI{\approx 50}{\micro \meter}$ in the $\hat{z}$ direction. This prevents excess of local increase of the atomic density, which could lead to heating and losses due to spin-flip collisions. Once the OCDT loading is complete, the two centers are brought closer by means of a uniform DC magnetic field ($B_{offset}\approx \SI{50}{\milli \gauss}$), produced by an additional pair of coils sharing its axis with the anti-Helmholtz coils.

After the RF is switched off, forced evaporation in the OCDT is performed by exponentially decreasing the total light power from $\SI{5}{\watt}$ to $\SI{0.04}{\watt}$ in $\SI{6}{\second}$; at the same time, the magnetic confinement is slightly reduced by ramping the MQT field gradient down from $\SI{20}{\gauss \centi \meter^{-1}}$ to $\SI{18}{\gauss \centi \meter}^{-1}$ (\textbf{e} in Fig.~\ref{fig:seq}). 

A holding time of $\SI{100}{\milli \second}$ is allocated for the final thermalization of the atoms and the on-set of  quantum degeneracy. The OCDT is stabilized to $\SI{0.04}{\watt}$ and a levitating magnetic field gradient of $\SI{18}{\gauss \centi \meter^{-1}}$ is maintained (\textbf{f} in Fig.~\ref{fig:seq}). A final `hold' phase of $\SI{10}{\milli \second}$ is then performed: the OCDT laser is switched off and atoms are kept in position against gravity only by the magnetic field gradient at $\SI{18}{\gauss \centi \meter^{-1}}$ (\textbf{g} in Fig.~\ref{fig:seq}). This allows hotter atoms to completely leave the volume monitored by the imaging system.

As described in the following, the onset of BEC is observed here. About $\SI{17}{\second}$ are required between the MOT phase and the onset of quantum degeneracy. At the end of the sequence, absorption images of the atom cloud are taken and the atoms temperature can be measured with free expansion time-of-flight measurements (\textbf{h} in Fig.~\ref{fig:seq}).

\subsection{Optical Crossed Dipole Trap Arrangement}\label{subsec:dt}
\begin{figure}[htbp]
\centering
\includegraphics[width=\linewidth]{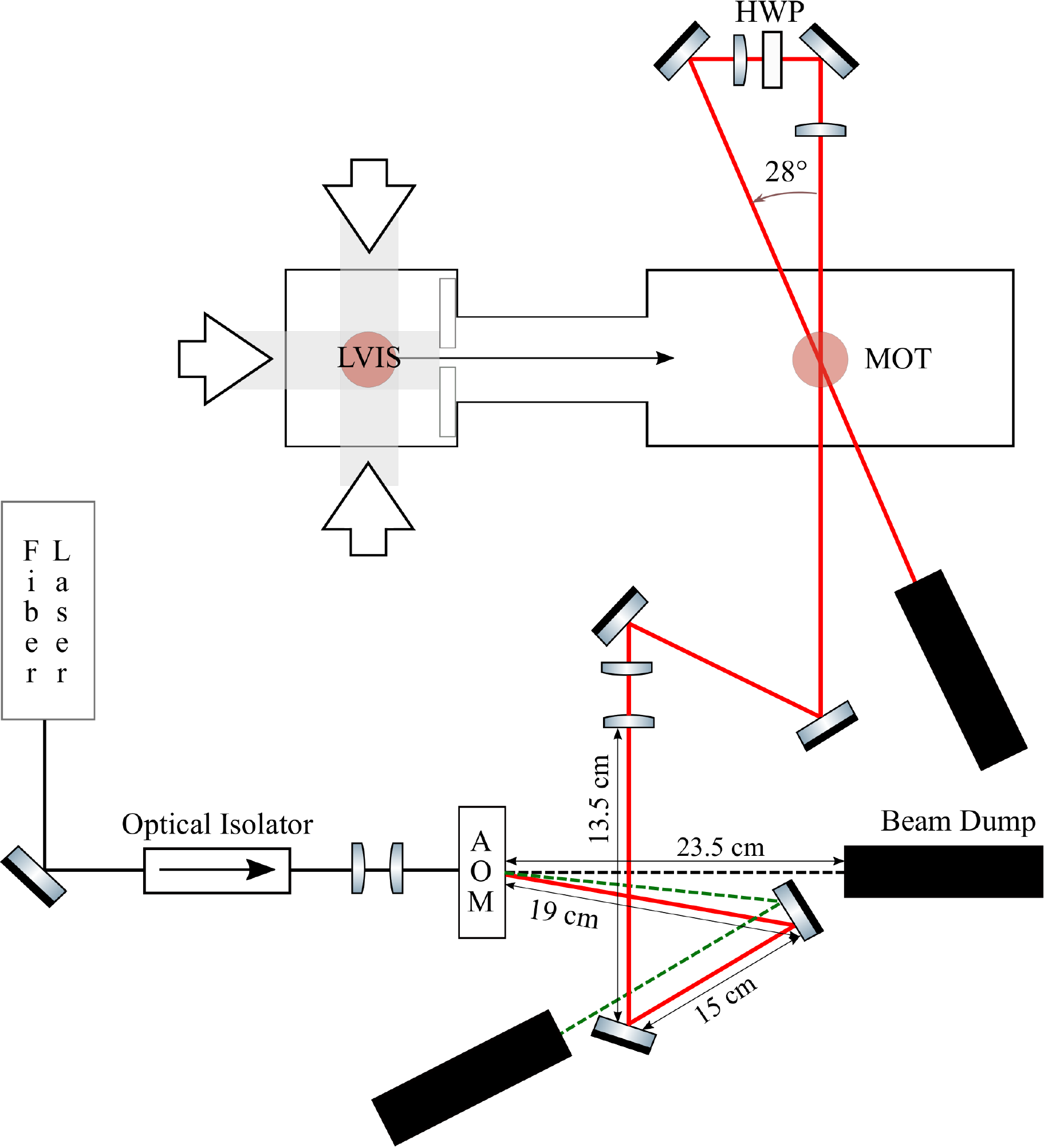}
\caption{Sketch of the OCDT setup. The continuous line represents the OCDT beams, whereas the dashed lines represent the diffraction orders generated by the AOM discarded along the optical path.}\label{fig:DTscheme}
\end{figure}

The experimental arrangement of the OCDT is sketched in Fig.~\ref{fig:DTscheme}. 

The OCDT beams are generated by a continuous wave single mode, linearly polarized Yb fiber laser (IPG ELR-20-LP-SF), emitting at $\SI{1070}{\nano \meter}$ with a maximum power of $\SI{20}{\watt}$. Back-reflections are prevented by a $\SI{33}{\decibel}$ free-space optical isolator (Thorlabs IO-5-1064-VHP).

Intensity stabilization and modulation, and fast beam switching are achieved by means of a free-space AOM (Isomet M1080-T80L), controlled by a Rohde $\&$ Schwarz SMY01 RF signal generator. In this way, the fiber laser can be always operated at full power, thus preventing thermal transients and delays. The laser beam crosses the atomic cloud and then is retroreflected with an angle of $\ang{28}$, as set by physical constraints of our setup. A half-wave plate (HWP) adjusts the relative orientation of the two beams' linear polarizations to perpendicular, in order to minimize interference. After adjustment of the HWP orientation, an increase up to $50\percent$ in the number of trapped atoms with respect to non-optimized conditions is observed. 

\begin{figure}[htbp]
\centering
\includegraphics[width=\linewidth]{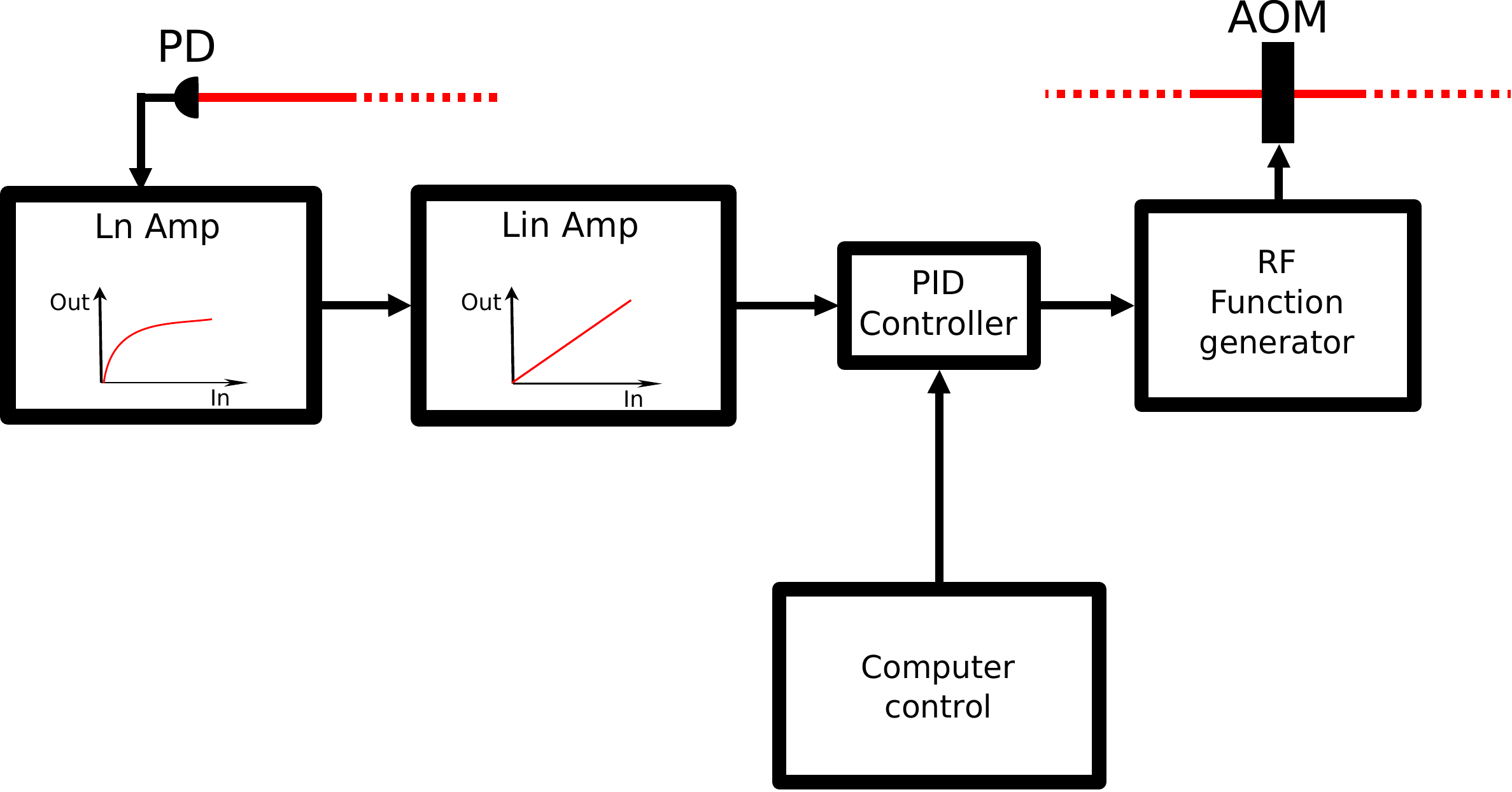}
\caption{Schematic of the intensity stabilisation circuit for the OCDT laser. PD: photodiode.}\label{fig:PowStab2}
\end{figure}

A schematic of the circuit for intensity stabilization is presented in Fig.~\ref{fig:PowStab2}. The intensity of the OCDT is controlled via a Labview computer interface. This sets the reference level (setpoint) for the proportional-integral-derivative (PID) controller (Stanford Research Systems SIM960), which acts on the amplitude modulation input of the RF function generator. The input for the feedback loop is obtained by monitoring a small fraction of the first OCDT beam with a Thorlabs DET36A/M photodiode. 

A chain of an AD8304 logarithmic amplifier and a Stanford Research Systems SIM910 JFET linear amplifier is installed (see Fig.~\ref{fig:PowStab2}), in order to maximize the dynamic range of the PID loop. In fact, the logarithmic amplifier allows a finer control in the critical low intensity regime, whereas the linear amplifier provides better performance in the high intensity regime. The output of the PID loop controls the RF supplied to the AOM, thus allowing robust control of the OCDT power.

\begin{figure*}[htbp]
\centering
\includegraphics[width=0.7\linewidth]{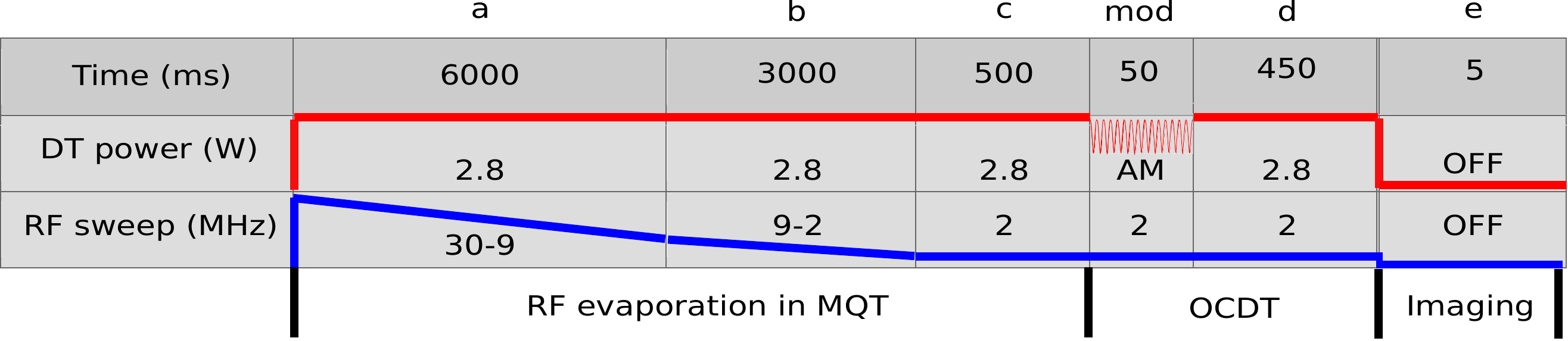}
\caption{Overview of the experimental sequence used for the OCDT frequency measurement. In \textbf{mod}, a modulation is applied to the OCDT laser intensity.}
\label{fig:FrqMeasSeq}
\end{figure*}

Fast and abrupt modulation of the RF amplitude injected in the AOM, however, can cause thermal transients in the modulator's crystal, with consequent power and positioning instability of the beams and reduced promptness of the system's response. This is prevented by driving the AOM with a constant RF power: in order to maintain the possibility of controlling the OCDT power, the acousto-optic modulator is hence driven by two sine waveforms at different frequencies \cite{aoms}. In our setup, the \textit{main} frequency, i.e.~that providing the beams used for the OCDT, is set to $\SI{95}{\mega \hertz}$, while the \textit{dummy} frequency is fixed at $\SI{65}{\mega \hertz}$, supplied by a Marconi Instruments 2024 signal generator. These values are chosen in order to maximize the angular separation of the corresponding diffraction orders, while complying with the AOM's operational bandwidth.

The two signals are summed by a Mini-Circuits ZMSC-2-1+ power splitter-combiner. The total RF power supplied to the AOM is kept constant by an analog circuit, which balances the amplitude modulations provided to the RF generators, keeping their sum constant. The implementation of this method resulted in a reduction of the horizontal beam drift by $>65\percent$ of its previous value and in the suppression of power transients when the OCDT is turned on.

The effective power available for the OCDT is $\SI{5}{\watt}$ in the first beam (along $\hat{x}$) and $\SI{3.5}{\watt}$ in the retro-reflected one. The corresponding beam waists at the trapping region are of $\SI{80}{\micro \meter}$ and $\SI{110}{\micro \meter}$, respectively, as measured by means of a Cinogy CinCam CCD-2302 beam profiler. Investigation with the beam profiler also confirmed the Gaussian profile (TEM$_{00}$) of the beams' wavefronts.

\subsection{Optical Crossed Dipole Trap Characterization}\label{subsec:dtchar}
As a test of the operational parameters of the OCDT,  an investigation of the harmonic trap frequencies is conducted. In particular, given the experimental parameters, a radial trap frequency of:

\begin{equation} 
\omega_{trap} = \sqrt{\dfrac{4U}{M_{Rb}w^2}}=870~Hz~\label{eqn:omegatrap}
\end{equation}

is predicted. Here, $U$ is the trap potential depth, $M_{Rb}$ the rubidium atom mass and $w$ the OCDT's beam waist. It is noteworthy that, given the fact that the Rayleigh range in the present case is $z_{R}\approx \SI{30}{\milli \meter}$, contributions from axial frequencies are negligible.


The radial trap frequency is measured by modulating the OCDT beam intensity during the loading phase. This causes losses due to parametric heating when the modulation matches the trap frequency. In this condition, by measuring the number of atoms loaded in the optical trap, the trapping frequencies can be measured \cite{trapfreq01, trapfreq02}. 

In order to implement this configuration, the output of a Keysight 33210A function generator is summed to the amplitude modulation signal produced by the OCDT intensity PID controller: the frequency of the function generator is varied and the corresponding population of the OCDT is measured. An overview of the experimental sequence used for the measurement is presented in Fig.~\ref{fig:FrqMeasSeq}.

\begin{figure}[htbp]
\centering
\includegraphics[width=\linewidth]{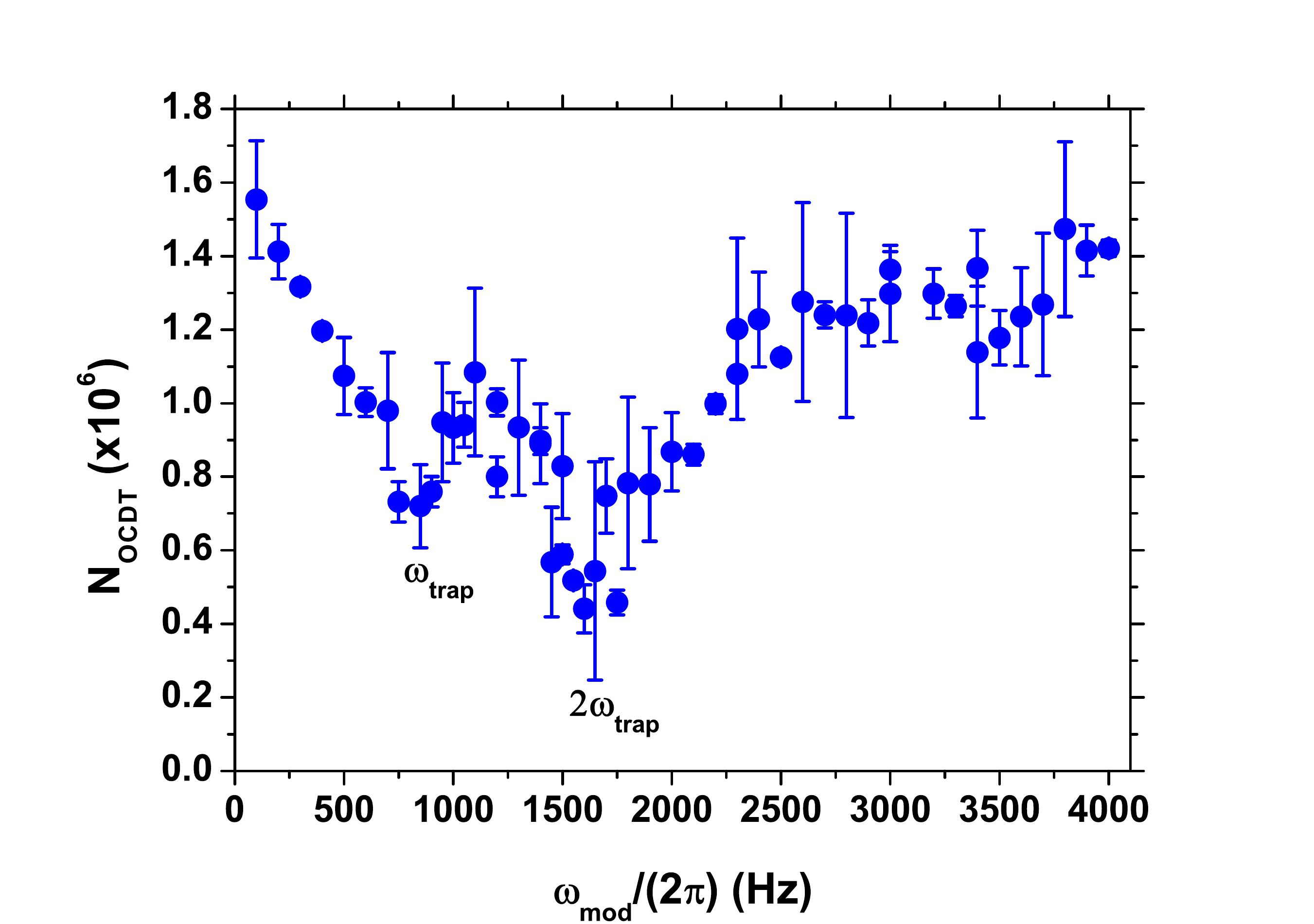}
\caption{OCDT radial frequency measurement: number of atoms in the OCDT ($N_{OCDT}$)  as a function of the laser intensity modulation frequency ($\omega_{mod}$). Errorbars indicate the standard deviation of the averaged values plotted in the graph.}\label{fig:TrapFreqMeas}
\end{figure}

The modulation was applied for $\SI{50}{\milli \second}$, with an amplitude corresponding to $50\percent$ of the laser intensity. The results are reported in Fig.~\ref{fig:TrapFreqMeas}.

In Fig.~\ref{fig:TrapFreqMeas}, resonant decreases in the number of trapped atoms at $\omega_{meas}\approx \SI{850 \pm 26 }{\hertz}$, where we have obtained an extra $\SI{\pm 10 }{\hertz}$ fluctuation among different measurements and fits, and $2\omega_{meas}$ are clearly visible. The measured quantity, as estimated by a Lorentzian fit, is roughly consistent with the predicted trap frequency ($\omega_{trap} = \SI{870}{\hertz}$).

\section{Experimental Results: Bose-Einstein Condensation}\label{sec:exp}
\begin{figure*}[htbp]
\centering
\includegraphics[width=\linewidth]{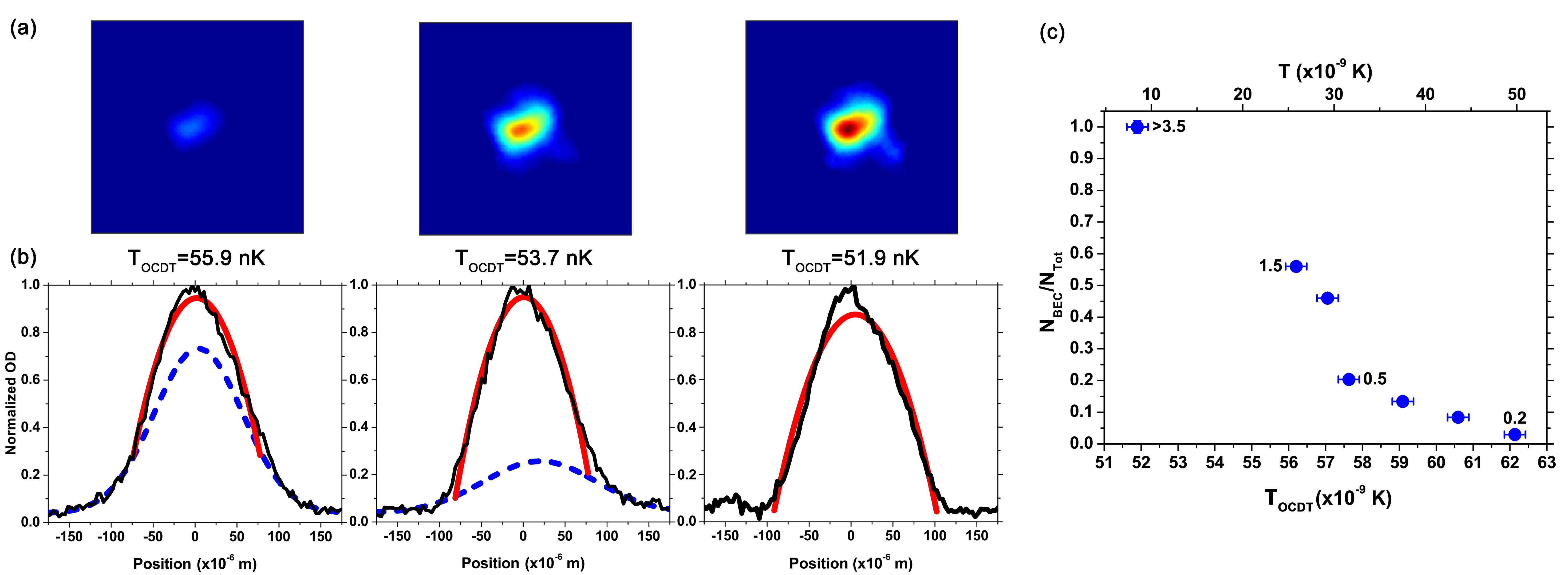}
\caption{\textbf{(a)} Absorption images of the ${}^{87}$Rb atomic cloud at different OCDT final depth, after $\SI{20}{\milli \second}$ free expansion. Color scale was chosen in order to highlight the BEC phase. Details are in the main text. The final atomic cloud temperature $T$ is $\SI{9.5\pm0.5}{\nano \kelvin}$. \textbf{(b)} Corresponding density profiles. Fits reveal the bimodal distribution of the atoms: dashed blue curves represent the Gaussian fit of the thermal component, while continuous red curves represent the parabolic fit of the BEC component. \textbf{(c)} Condensed fraction ($N_{BEC}/N_{Tot}$) as a function of the OCDT final depth, after $\SI{20}{\milli \second}$ free expansion. Numbers next to experimental points label the corresponding phase space density (Eq.~\ref{eqn:psd}), extrapolated for $\SI{0}{\milli \second}$ free expansion.}\label{fig:BECfit}
\end{figure*}

Forced evaporation in the OCDT causes the phase transition from a thermal gas to a BEC. As the bimodal distribution of the atomic cloud confirms, this is observed at about $T_{OCDT}\approx \SI{60}{\nano \kelvin}$, where $T_{OCDT}$ is a measure of the OCDT depth. Eventually, a pure condensate of $5 \times 10^4$ atoms is consistently obtained every $\SI{50}{\second}$, at an average temperature of  about $T\approx \SI{10}{\nano \kelvin}$. As anticipated, the actual evaporation to quantum degeneracy takes $\SI{17}{\second}$. It is worth underlining that $T_{OCDT}$ represents the theoretical trap depth and, therefore, is considered here an exact parameter. $T$, instead, is the actual atomic temperature, measured by free expansion.

In Fig.~\ref{fig:BECfit}.(a), examples of absorption images taken during the final evaporation are shown. The corresponding atomic density distribution is investigated in \ref{fig:BECfit}.(b), where the density profile as measured from absorption imaging is fit by a Gaussian curve for the thermal cloud and a quadratic curve for the BEC phase. The progressive increase of the BEC component, up to $100\percent$ at $T_{OCDT}=\SI{51.9}{\nano \kelvin}$ (corresponding to $T=\SI{9.5\pm0.5}{\nano \kelvin}$), is highlighted by the color scale: in (a), colormaps are normalized by taking into consideration the amplitude of the quadratic density peak produced by the condensed atoms. In particular, the minimum level of the color scale is set by the maximum of the Gauss fit curve, thus `isolating' the BEC contribution from the overall atomic density.

Fig.~\ref{fig:BECfit} also further demonstrates the high level of control on the OCDT achieved thanks to the stabilization system described above: the possibility of controlling the temperature of the cloud within $\sim \SI{0.1}{\nano \kelvin}$ is systematically confirmed.

The onset of the BEC is confirmed by calculating the condensed fraction $N_{BEC}/N_{Tot}$ as a function of the final trap depth, as shown in Fig.~\ref{fig:BECfit}.(c). The fraction is computed as the ratio between the area enclosed by the parabolic fit of the density distribution (BEC phase) and that of the entire atomic cloud, corresponding to the total area enclosed by the Gaussian curve (thermal phase) and the BEC phase. The first evidence of quantum degeneracy is found at $T_{OCDT}\sim \SI{62}{\nano \kelvin}$. This corresponds to an OCDT final power of $\SI{40}{\milli \watt}$.

As a further diagnostic of the atomic sample, the phase space density $PSD$ can be introduced:

\begin{equation}
PSD=n \lambda_{dB}^{3}~, \label{eqn:psd}
\end{equation}

where $n$ is the density of the atomic cloud and $\lambda_{dB}=\sqrt{2 \pi \hbar^{2}/(M_{Rb}k_{B}T)}$ is the de Broglie wavelength associated to the trapped atoms. $PSD$ values corresponding to the experimental points are reported in Fig.~\ref{fig:BECfit}.(c). It is noteworthy that $PSD\geq10$ have been measured. However, given the optical thickness of the atomic sample in these conditions, the experimental measure of the spatial density may be not trustworthy. Consequently, these measurements were not taken into consideration here. The maximum $PSD$ is therefore in excess of $3.5$, as measured with $N_{BEC}/N_{Tot}=1$ at $T_{OCDT}=\SI{51.9}{\nano \kelvin}$. The role of the final `hold' of $\SI{10}{\milli \second}$ has revealed of major importance for  the background thermal atoms to leave the experimental volume. They would have been in fact detected as an offset in the density, which can affect the $N_{BEC}/N_{Tot}$ value.

Moreover, by analyzing data shown in Fig.~\ref{fig:BECfit} with the scale law for interacting atoms in a BEC in a harmonic trap \cite{foot}: 

\begin{equation}
\dfrac{N_{BEC}}{N_{Tot}}=1-\left(\dfrac{T}{T_{c}}\right)^{3}, \label{eqn:becfraction}
\end{equation}

where $T_{c}$ is the BEC critical temperature, a critical temperature of $T_{c}^{exp}\approx \SI{57.5\pm0.6}{\nano \kelvin}$ can be estimated. However, deviations from Eq.~\ref{eqn:becfraction}, mostly due to further residual interactions and the finite size of the harmonic confining potential of the OCDT, prevent a more solid estimation.

\begin{figure}[htbp]
\centering
\includegraphics[width=\linewidth]{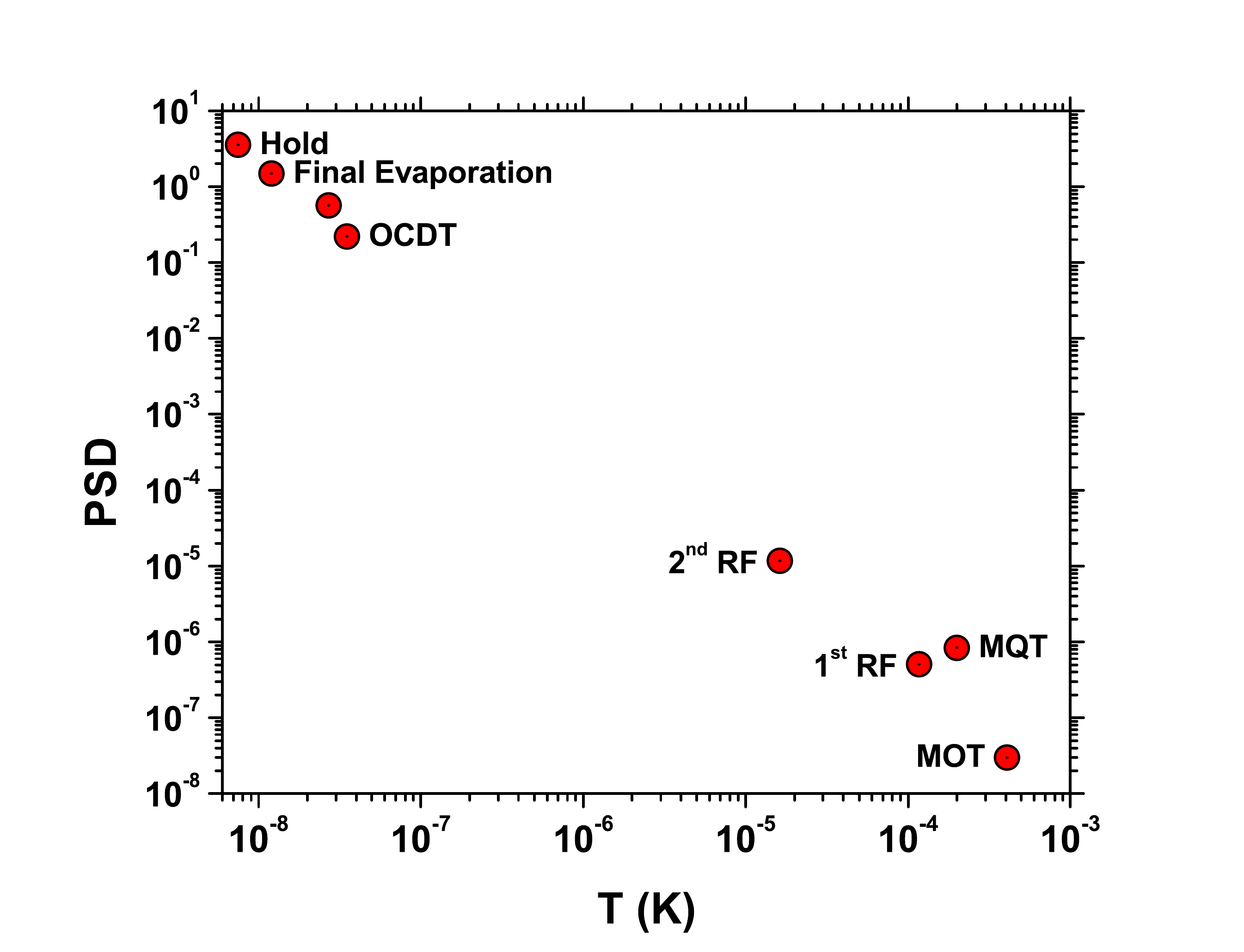}
\caption{Evolution of $PSD$ as function of the temperature of the atomic cloud throughout the experimental sequence. Error bars are concealed by the markers representing experimental points.}\label{fig:PSDvsTemperature}
\end{figure}

Finally, in order to characterize the experimental sequence as a whole, in Fig.~\ref{fig:PSDvsTemperature} the evolution of the $PSD$, as defined in Eq.~\ref{eqn:psd}, is plotted as a function of the atomic temperature.

From a starting value of $PSD \approx 10^{-8}$ in the MOT in the science chamber, the phase space density increases to $PSD \geq 3.5$ in $\SI{17}{\second}$, corresponding to a temperature decrease from $\SI{400}{\micro \kelvin}$ to $\SI{10}{\nano \kelvin}$. Remarkably, the largest increase is obtained with the OCDT: optical forced evaporation increases the $PSD$ from $10^{-5}$ to $1$ in only $\SI{6.1}{\second}$.

\section{Conclusions}
We have described an experimental setup for reliable generation of ${}^{87}$Rb BECs from thermal vapor. Our setup employs an isotopically selected, pre-cooled, collimated beam of ${}^{87}$Rb atoms obtained from an LVIS source. Technical details of the MOT compression and sub-Doppler cooling are described, as well as the loading of the MQT, where preliminary evaporation by two-stages RF coupling to untrapped states is performed. Overcompensation of the background DC magnetic fields allows fast switching of the uniform magnetic field required for optical pumping, which can be thus reduced to only $\SI{0.2}{\milli \second}$, in order to avoid unnecessary heating of the atomic cloud. After loading in a hybrid optical crossed dipole trap, a single stage forced evaporation is performed by exponentially decreasing the OCDT intensity. Fine and robust control of the OCDT beams' intensity and position is achieved by dual-frequency driving of the AOM used for fast switching of the trap beams. This allows the elimination of thermal transients and power and position instability. A chain of logarithmic and linear amplifiers allows to increase the dynamic range of the PID loop stabilizing the OCDT intensity.

Our system is capable of producing a pure BEC of $5\times 10^{4}$ ${}^{87}$Rb atoms at $T=\SI{9.5\pm0.5}{\nano \kelvin}$ every $\SI{50}{\second}$, with an evaporation time towards quantum degeneracy in the $|F=2, m_{F}=+2\rangle$ state of $\SI{17}{\second}$. The setup can be easily reconfigured to different experimental sequences via Labview and can be easily adapted to trap atoms in the more frequently explored $|F=1, m_{F}=-1\rangle$ state, by implementing a temporal dark MOT prior to the gradient cooling,  and suitably adapting the optical pumping pulse and the QMT trapping potential.

\bibliography{BECpaper}

\begin{acknowledgments}
Financial support from the Leverhulme Trust (Grant No.~RPG 2012 809) is acknowledged.

The authors would like to thank Dr Alessia Burchianti (INO and LENS - University of Florence), Dr Nicole Fabbri (LENS - University of Florence), Dr Chiara Fort (LENS - University of Florence), Dr Matt Jones (Nottingham University) and Dr Anna Marchant (Durham University) for the fruitful discussions and precious suggestions.
\end{acknowledgments}

\end{document}